\documentclass[aps,11pt,onecolumn,preprintnumbers,amsmath,amssymb]{revtex4}

\usepackage[english]{babel}
\usepackage{graphicx}
\usepackage{color}

\begin{document}

\newcommand{\unit}[1]{\:\mathrm{#1}}            
\newcommand{\To}{\mathrm{T_0}}
\newcommand{\Tp}{\mathrm{T_+}}
\newcommand{\Tm}{\mathrm{T_-}}
\newcommand{\EST}{E_{\mathrm{ST}}}
\newcommand{\Rp}{\mathrm{R_{+}}}
\newcommand{\Rm}{\mathrm{R_{-}}}
\newcommand{\Rpp}{\mathrm{R_{++}}}
\newcommand{\Rmm}{\mathrm{R_{--}}}
\newcommand{\ddensity}[2]{\rho_{#1\,#2,#1\,#2}} 
\newcommand{\ket}[1]{\left| #1 \right>} 
\newcommand{\bra}[1]{\left< #1 \right|} 

\title{Optically Active Quantum Dots in Monolayer WSe$_2$}
\author{Ajit Srivastava$^{1*}$}
\author{Meinrad Sidler$^{1*}$}
\author{Adrien V. Allain$^2$}
\author{Dominik S. Lembke$^2$}
\author{Andras Kis$^2$}
\author{A. Imamo\u{g}lu$^1$}
\affiliation{$^1$Institute of Quantum Electronics, ETH Zurich, CH-8093
Zurich, Switzerland}
\affiliation{$^2$Electrical Engineering Institute, Ecole Polytechnique Federale de Lausanne (EPFL), CH-1015
Zurich, Switzerland.}

\maketitle

*Correspondence to: imamoglu@phys.ethz.ch, sriva@phys.ethz.ch

\textbf{Semiconductor quantum dots have emerged as promising
candidates for  implementation of quantum information processing
since they allow for a quantum interface between stationary spin
qubits and propagating single
photons~\cite{Divincenzo00,Imamoglu99,Gao12}. In the meanwhile,
transition metal dichalcogenide (TMD) monolayers have moved to the
forefront of solid-state research due to their unique band structure
featuring a large band gap with degenerate valleys and non-zero
Berry curvature~\cite{Xu14}. Here we report the observation of
quantum dots in monolayer tungsten-diselenide with an energy that is
20 to 100 meV lower than that of two dimensional excitons. Photon
antibunching in second-order photon correlations unequivocally
demonstrates the zero-dimensional anharmonic nature of these quantum
emitters. The strong anisotropic magnetic response of the spatially
localized emission peaks strongly indicates that radiative
recombination stems from localized excitons that inherit their
electronic properties from the host TMD. The large $\sim 1$~meV
zero-field splitting shows that the quantum dots have singlet ground
states and an anisotropic confinement most likely induced by
impurities or defects in the host TMD. Electrical control in van der
Waals heterostructures~\cite{Lee14} and robust spin-valley degree of
freedom~\cite{Xiao12} render TMD quantum dots promising for quantum
information processing.}\\

\textbf{Main Text} \\
Advances in semiconductor based quantum information processing  have
been made on two disjoint fronts: while optically active
self-assembled quantum dots (QD) with deep electron and hole
confinement allow for the realization of highly efficient
single-photon sources~\cite{Dousse10}, all-optical
manipulation of confined spins~\cite{Berezovsky08,Press08} and a spin-photon
quantum interface~\cite{Gao12,DeGreve12}, the random nature of their
growth seems to be the biggest hinderance towards their use in
scalable quantum information processing. In contrast,
electrically-defined single~\cite{Elzerman04} or double
QDs~\cite{Petta05} hosting one or two excess electrons have been
shown to exhibit long spin coherence times together with a clear
path towards integrated scalable devices. However, weaker
confinement has precluded the possibility to reliably transfer
quantum information from spins to photons in these systems. QDs in
monolayer TMDs have the potential to combine the desirable features
of both optically active and electrically-defined QDs: while we
report tungsten-diselenide (WSe$_2$) QDs that appear due to
uncontrolled impurity or defect induced traps, the two dimensional
(2D) nature of these materials makes it easier to electrically
control the local potentials on a scale of few tens of nanometers.
More importantly, strong electron-hole binding in TMDs suggest that
it would be possible to obtain a quantized optical excitation
spectrum due to trapping of excitons or trions in large electric
field gradients induced by external gates~\cite{Wilson-Rae12}.

The samples studied in our experiments were obtained by mechanical
exfoliation from WSe$_2$ synthetic crystals onto heavily doped
silicon substrates with 285 nm SiO$_2$ layer on
top~\cite{Novoselov05,Radisavljevic11}. Monolayer flakes were
identified using their optical contrast. Polarisation-resolved
photoluminescence (PL) and resonant white-light reflection
spectroscopy were performed using a home-built confocal microscope
setup placed in a liquid helium bath cryostat. The sample
temperature was 4.2 K and the excitation source was a helium-neon
(HeNe) laser at 632.8 nm or a tunable continuous-wave (cw)
Ti:Sapphire laser. The spot size for excitation wavelength was $\sim
2 \mu$m.

Figure 1a (left panel) shows photoluminescence (PL) spectra of a monolayer flake (flake0)
of WSe$_2$ not showing sharp emission peaks at two different
excitation laser powers ($P_{exc}$). The spectra clearly shows two
high energy peaks: exciton (X$^{0}$) emission at 708 nm, charged
exciton (trion) emission (X$^{-}$) at around 722 nm at high
$P_{exc}$. These assignments are in accordance with previous reports
of PL from monolayer WSe2~\cite{Jones13}. The third feature of the
spectra is the broad emission to the lower energy side of the trion
peak, that has been previously attributed to impurity/defect trapped
excitons in WSe$_2$ and other TMDs~\cite{Tongay13}. At low
$P_{exc}$, X$^0$ and X$^-$ peaks are too weak to be detected while
the impurity band peak can still be seen, suggesting starkly
different power-dependence of the impurity band as compared to X$^0$
and X$^-$ peaks. A detailed power dependence of X$^0$, X$^-$ and the
impurity band shown in Figure 1a (right panel) confirms this claim
- while the integrated PL intensity of X$^0$ and X$^-$ peaks shows
linear dependence up to the highest  ($P_{exc}\sim$ 320 $\mu$W),
that of the impurity band exhibits saturation behaviour (sub-linear
dependence) even at powers as low as $\sim$ 100 nW. We do not
observe any broadening of the X$^0$ and X$^-$ peak even at the
highest $P_{exc}$ while the spectral features of the impurity band
change significantly with incident laser power (see Supplementary Fig.~S1).


This type of saturation behavior
at relatively low powers compared to free excitons is indeed
consistent with PL that stems from defects and impurities which act
as two-level emitters. It is well known from optical investigation of
quantum well structures in III-V semiconductors that in low quality
samples with a high defect or impurity density, the PL is dominated
by red-shifted emission from these localized state. This is a
consequence of the fact that optically generated excitons relax by
phonon emission into the localized states, quenching the delocalized
exciton emission.

Figure 1b shows PL spectra for another WSe$_2$ flake (flake1)
where the impurity band is comprised of sharp emission peaks which
become prominent at lower $P_{exc}$ and finally for $P_{exc} \le 1
\mu$W, only sharp peaks are visible while the X$^0$ peak becomes too
weak to be detected.  Figure 1c (right panel) shows a spatial map of integrated
PL intensity for X$^0$ (area enclosed by the black contour in Figure 1c) and two different sharp emission peaks (left panel) on
flake1 obtained by raster scanning the sample with piezoelectric
positioners. The sharp emission peaks are seen to be spatially
localised such that PL spectra at energies lower than X$^-$ emission
appear strikingly different at different locations on the flake. High resolution PL spectrum of the two sharp peaks, labeled QD1F1 and QD2F1 are
obtained with $P_{exc} <$ 1 $\mu$W and exhibit extremely narrow linewidths
of less than 120 $\mu$eV, possibly limited by spectral diffusion of
the emission peak as discussed below. We rule out Raman scattering
as the origin of the sharp features since changing the energy of the
incident laser does not result in any corresponding shift in the
emission energy. It is interesting to note that we
observe strong PL from a region (coloured red in Figure 1c) where the exciton emission is more
than an order of magnitude weaker than its maximum value. We also measured
differential reflectance of the flake using a broadband white light
source which shows a clear peak only for X$^0$~\cite{valleyZeeman}.

To prove that the sharp emission peaks originate from localized
excitons with an anharmonic spectrum, we measure the photon
correlation function $g^{2}(\tau)$ using a Hanbury-Brown-Twiss (HBT)
setup with two single photon counting avalanche photodiodes (APDs). Since $g^{2}(\tau=0)$
gives the likelihood for detecting two photons simultaneously, an
ideal single zero-dimensional (0D) emitter with uncorrelated
background contribution gives $g^{2}(0) = 0$. In general, a measured
value of $g^{2}(0) < 0.5$ unequivocally proves that the source of
emission originates predominantly from a single anharmonic emitter
-- a quantum dot (QD). Figure 2a shows the $g^{2}(\tau)$
measurement results for QD1F1 and QD2F1, yielding $g^{2}(0) = 0.18
\pm 0.02$ and $g^{2}(0) = 0.20 \pm 0.02$, respectively. We therefore
conclude that the sharp peaks are indeed associated with QD PL.
Figure 2b shows the PL lifetime  of QDs QD1F1 and QD3F1 on flake1
measured by exciting with a $\sim$ 5 ps pulsed Ti:Sapphire laser
tuned into resonance with  X$^0$ transition and sending the
spectrally filtered output around the QD wavelength to a
single-photon-counting APD with a timing resolution of $\sim$ 350
ps. The measured long lifetime of few nanoseconds is consistent with
behavior shown by typical semiconductor quantum dots. In contrast,
the exciton lifetime in MoS$_2$ is reported to be about 4
ps~\cite{Lagarde14} and is determined predominantly by non-radiative
decay.

It is well known that 0D emitters in solids typically exhibit
spectral diffusion~\cite{Frantsuzov08}, blinking and in some cases
photo-bleaching. To determine if WSe$_2$ QDs also exhibit these
properties that strongly limit their applications, we record the
single photon detection events for QD1F1 using a single-photon
counting APD with a dead time of $\sim$ 200 ns at a total photon
counting rate of $\sim$ 13 kHz. We then use this data to obtain the
waiting time distribution W($\tau$) for $\tau \ge 10 \mu$s. Figure
3a shows that W($\tau$) can be fitted by a single exponential where
the decay time is given by $\tau_{det} = (\Gamma \eta)^{-1}$ $\sim$
77 $\mu$s, in excellent agreement with the measured photon counting
rate. Here $\Gamma$ is the spontaneous emission rate and $\eta$ is
the detection efficiency. This observation shows that there is no
blinking for timescales longer than $\tau_{det}$: in the event of
intensity intermittency or blinking one expects the intensity to
exhibit bright and dark states between which the system switches
randomly. If this switching takes place on a time-scale
$\tau_{blink} > \tau_{det}$, then W($\tau$) should have an
additional decay on a timescale determined by
$\tau_{blink}$~\cite{Delteil14}. The absence of bunching on
timescales $\le$ 1 $\mu$s in $g^{2}(\tau)$ measurements (not shown)
in turn rule out blinking on such short timescales. Finally, we
emphasize that blinking statistics in most QDs is known from prior
studies to be non-Poissonian, exhibiting power-law tails in waiting
time distribution~\cite{Frantsuzov08}. In the absence of any such
deviation from an exponential decay, we tentatively conclude that
there is no significant blinking in the QDs studied here.

We emphasize that there is no obvious degradation of PL intensity up
to hundreds of $\mu$Ws of incident HeNe laser power, way above the
linear response regime of QDs. The QD PL also survived several
cycles of warm-up to room temperature and cool-down to 4.2 K; these
observations allow us to conclude that the QDs do not photo-bleach.
However, spectral diffusion of the emission peak with excursion
range of a few hundred $\mu$eV is observed in almost all QDs but
with varying strength. Most QDs exhibit small spectral diffusion
with range $\pm 200 \mu$eV which cannot be resolved using  a
low-resolution spectrometer (Fig. 3b). On the other hand, we
observed that some QDs exhibited large spectral diffusion of range
$\pm 0.5$ meV (Fig.~3c); such strong spectral diffusion, probably
arising from charge fluctuations in the environment of the QDs,
could last up to minutes before the peak energy is stable again.




Having established that the sharp emission peaks stem from optically
active QDs, we focus on demonstrating that the QDs inherit their
electronic properties, such as the valley degree of freedom, from
WSe$_2$.  A priori, it is not clear whether the QD confinement
strongly mixes the two valleys. Below we provide evidence that there
is no valley hybridization in TMD QDs studied here.

A first test in this regard is provided by
photoluminescence-excitation (PLE) measurement where we tune an
excitation laser across the free exciton and trion peaks and monitor
the PL intensity of QDs. As we demonstrate in Fig.~S2 (see
Supplementary), we observe an overall enhancement in PL for most QDs
suggesting efficient relaxation from free excitons to QD states.
Remarkably, for all QDs on which PLE was performed, there is a sharp
cut-off for excitation laser energy below which the PL signal drops
drastically. As PLE probes the absorption spectra of the emitter,
such a cut-off in laser energy could mark the onset of continuum
states of the QD below which only sharp discrete states could exist.

To demonstrate the existence of a valley degree of freedom for the
observed QDs, we perform polarisation-resolved magneto-optical
spectroscopy with magnetic field (B-field) both perpendicular
(Faraday geometry) and parallel (Voigt geometry) to the flake. It
has been recently reported that only in the presence of a magnetic
field perpendicular to the TMD flake, exciton and trion emission
peaks split into two circularly polarised peaks of opposite
helicities due to the lifting of valley
degeneracy~\cite{valleyZeeman}. This strongly anisotropic magnetic
response was attributed to two-dimensional electronic structure of
TMDs. Magnetic field dependence of PL obtained by a linearly
polarized excitation laser and detected in circularly-polarized
basis in Faraday geometry is shown in Figure 4a (left panel): two
split peaks are observed even at zero B-field with the splitting
increasing with B-field. Most notably, no clear splitting is
observed in the Voigt geometry just like in the case of exciton and
trion (Fig.4a, right panel). The split peaks are circularly
polarized with opposite helicities at finite B-field which reverse
upon reversing the direction of B-field confirming that the peaks
arise from the same QD (Fig. 4b). These observations strongly
suggest that the QDs inherit their electronic structure from the
host TMD and are very likely to be excitons trapped in shallow
defect or impurity potentials. Our findings also lend support to the recent prediction that valley hybridization
is absent in TMD QDs, preserving the valley physics of 2D bulk~\cite{Liu14QD}.

As the splitting between the peaks increases at higher fields, the
higher energy peak decreases in intensity. This is expected if there
is efficient thermalization between the two levels before
recombination as PL signal from a level is proportional to its
occupancy. Although in most zero-dimensional systems, such a
thermalization is highly inefficient, finite carrier density in TMDs
could, in principle, enhance thermalization.

The observed zero-field splitting of about 700 $\mu$eV possibly
originates from electron-hole exchange interaction of a neutral
exciton trapped in an asymmetric confining potential, as is commonly
observed in self-assembled InAs/GaAs quantum dots~\cite{Gammon96}.
It could also arise if the exciton is bound to an ionized donor as
no electron-hole exchange interaction is present when the exciton is
bound to a neutral donor with a spin-1/2 ground-state. It is
noteworthy that this zero-field splitting in TMD QDs is almost 50
times larger than that in InAs/GaAs self-assembled quantum
dots~\cite{Gammon96}, consistent with the strong Coulomb
interactions in monolayer TMDs.

The splitting between the two peaks is plotted against B-field in
Figure 4c and fit to a hyperbolic dispersion $E(B) = \sqrt{\mu^2 B^2
+ \delta_0^2}$ expected for exchange-mixed circularly polarized
resonances. We obtain $|\delta_0|$ between 650 to 850 $\mu$eV and
$\mu$ between 7.5 to 10.9 $\mu_B$ (Bohr magneton) after fitting
splitting data for several different QDs in two different flakes
(see Fig.~S3, supplementary). This measured magnetic moment of QD
exciton is significantly larger than that of a free exciton, $\sim$
4 $\mu_B$ and a trion $\sim$ 5.5 to 6 $\mu_B$ in
WSe$_2$~\cite{valleyZeeman}.


An earlier study on defect activated PL in TMDs concluded that the
impurity band emission arises from excitons trapped in anion
vacancies~\cite{Tongay13} which could also be the origin of QDs
studied here. In addition, recent high-resolution transmission
electron microscopy (TEM) studies have revealed the presence of line
defects and island-like domains within single MoS$_2$
layer~\cite{Enyashin13}; it is plausible that such defects lead to
exciton localization. Finally we remark that we observed strong
emission from individual QDs in two flakes while most other flakes
showed behaviour similar to flake0 presumably due to a much higher
density of QDs leading to a formation of impurity band.

QDs in TMDs are likely to have very favorable properties for
applications in quantum information processing~\cite{Kormanyos14}.
Since we have established that quantum confinement does not lead to
valley hybridization, a qubit defined using the two lowest energy
states of an excess QD electron could be classified as a spin-valley
qubit. Given that the spin and valley degrees of freedom are
strongly correlated due to large spin-orbit interaction, any
decoherence mechanism that couples only to spin or to valley will
remain ineffective. Unlike GaAs, the dominant isotopes of most TMDs
are nuclear-spin free, rendering intrinsic hyperfine decoherence to
be weak. An open fundamental question motivated by our work is the
origin of the anomalously large g-factor of TMD QDs. Understanding
the nature of quantum confinement would provide insights for
engineering QDs within a monolayer.
\\

%


{}

\vspace{1 cm}

\textbf{Acknowledgments} This work is supported by NCCR Quantum Science and
Technology (NCCR QSIT), research instrument of the Swiss National Science Foundation (SNSF)\\

%
%

\newpage

{\bf Figure 1: Photoluminescence of monolayer WSe$_2$ flakes}  {\bf a,} (Left panel) Low temperature (4.2 K) photoluminescence (PL) spectra of a monolayer flake (flake0) not showing sharp peaks in emission at two different powers of incident laser (Helium-neon, 632.8 nm). Vastly different power dependence is observed for the exciton (X$^0$), trion (X$^-$) peaks and the broad feature (impurity band) at longer wavelengths. (Right panel) Detailed laser power dependence shows linear behaviour of the integrated PL intensity for X$^0$ (blue circles) and X$^-$ (green diamonds) peaks whereas the impurity band (red squares) PL has sub-linear dependence. {\bf b,} Low temperature PL spectra of a monolayer flake (flake1) showing sharp emission lines at high (low) laser power is plotted in the left (right) panel. X$^0$ peak is too weak to be detected at low power whereas the sharp emission peaks show saturation behaviour similar to the impurity band PL in Figure 1a. {\bf c,} (Right panel) A spatial map of PL from flake1 showing localised emission of the sharp peaks. The green (blue) region is the spatial extent of the PL peak labelled QD1F1 (QD2F1) in relation to the region where the X$^0$ PL reduces to half of its maximum value (area enclosed by the black contour). Red region depicts the spatial extent of sharp peak located close to the edge of the flake which has reduced overlap with X$^0$ region. (Left panel) High resolution PL spectra of QD1F1 and QD1F2 showing extremely narrow linewidths.
\\

{\bf Figure 2: Photon correlations and photoluminescence lifetimes of quantum dots} {\bf a,} Second order photon correlation function (g$^2$($\tau$)) of photoluminescence (PL) measured using Hanbury-Brown-Twiss setup shows pronounced dip (antibunching) at zero time delay  for the emission lines QD1F1 (left panel) and QD4F1 (right panel), confirming that they originate from zero-dimensional emitters - quantum dots. {\bf b,} Time-resolved PL of QD1F1 (left panel) and QD3F1 (right panel) measured using pulsed Ti:Sapphire laser of $\sim$ 5 ps shows long lifetime of the excited state typical of zero-dimensional emitters.
\\

{\bf Figure 3: Stability of emission and spectral wandering of quantum dots.} {\bf a,} Waiting time distribution of photoluminescence (PL) from QD1F1 as function of time delay between successive single photon detection events shows single exponential decay, indicating absence of intensity intermittency on timescales larger than the average time for photon detection, $\tau_{det}$. {\bf b,} (Left panel) Time-trace of PL emission of a typical quantum dot (QD) measured with low spectral resolution shows stable peak position. (Right panel) High resolution PL time-trace of another QD showing spectral wandering with a range of 1 meV. The synchronized wandering of the two peaks, strongly suggests that they arise from the same QD and are associated with electron-hole exchange split exciton transitions.
\\

{\bf Figure 4: Magnetic field dependence of quantum dot photoluminescence.} {\bf a,} (Left panel) Polarisation resolved magnetic field ($B$) dependence of photoluminescence (PL) from a quantum dot (QD1F2) in perpendicular $B$ (Faraday geometry) shows splitting of the two peaks with increasing $B$. The PL is detected in a circularly polarised basis ($\sigma_1$). Even though the high energy peak should be dominant at higher $B$ because of its emission helicity being that of the detection basis, efficient thermal relaxation causes the low energy peak to be stronger. In the opposite detection basis ($\sigma_2$), (see Fig.~4b), the low energy peak is even stronger for $B$ $>$ 0. (Right panel) $B$ parallel to the sample (Voigt geometry) does not show any measurable splitting. {\bf b,} The helicity of polarisation of the two split peaks of quantum dot QD2F1 denoted by red and blue traces switches sign upon reversal of the direction of $B$. {\bf c,} Extracted splitting between the two peaks in Faraday geometry of QD2F1 (left panel) and QD1F2 (right panel) as a function of $B$ is fitted with a hyperbolic function (see text) to extract the magnetic moment and the zero-field exchange splitting.
\\

\clearpage

\begin{figure}
\includegraphics[scale=0.8]{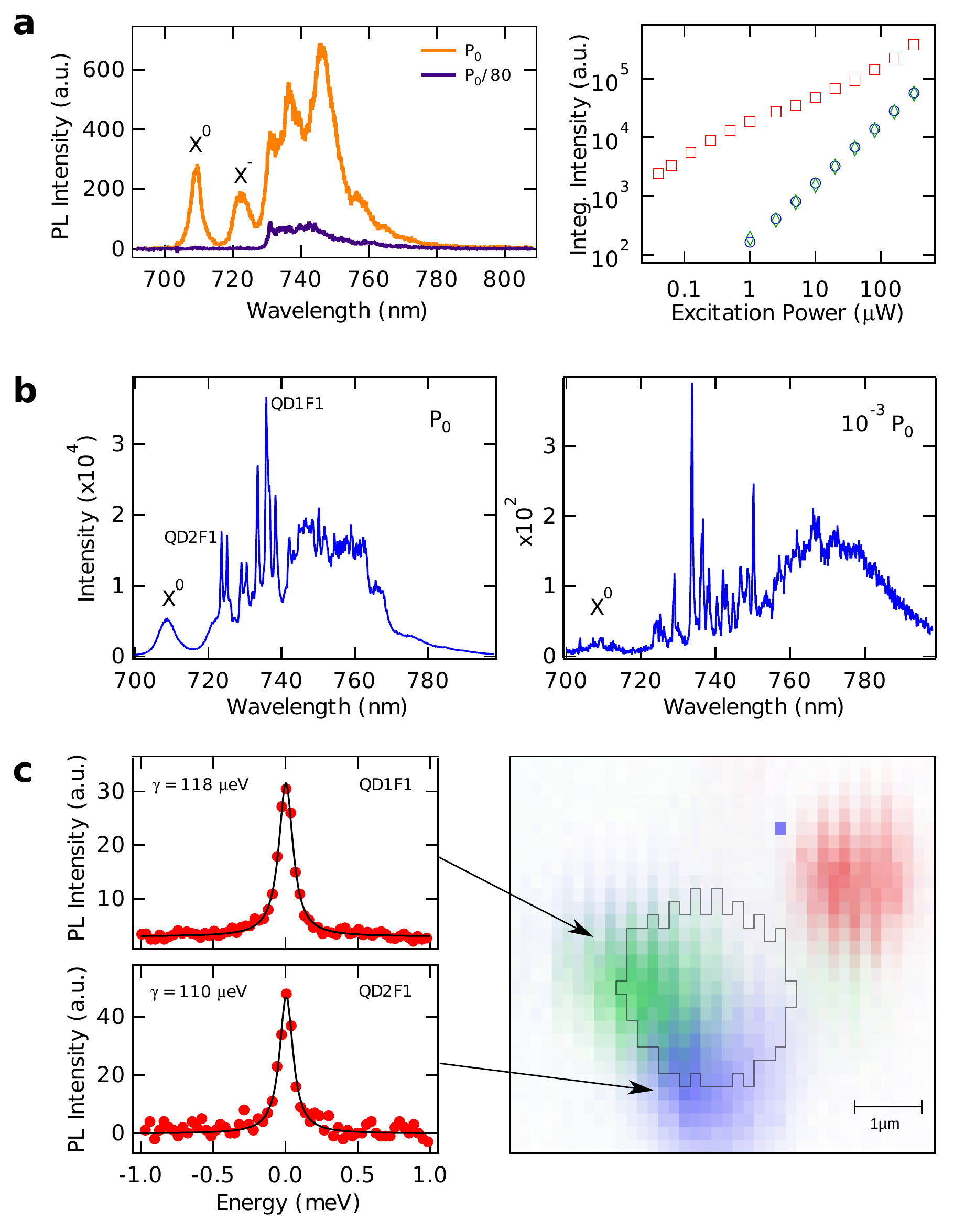}
\end{figure}

\clearpage

\begin{figure}
\includegraphics[scale=0.8]{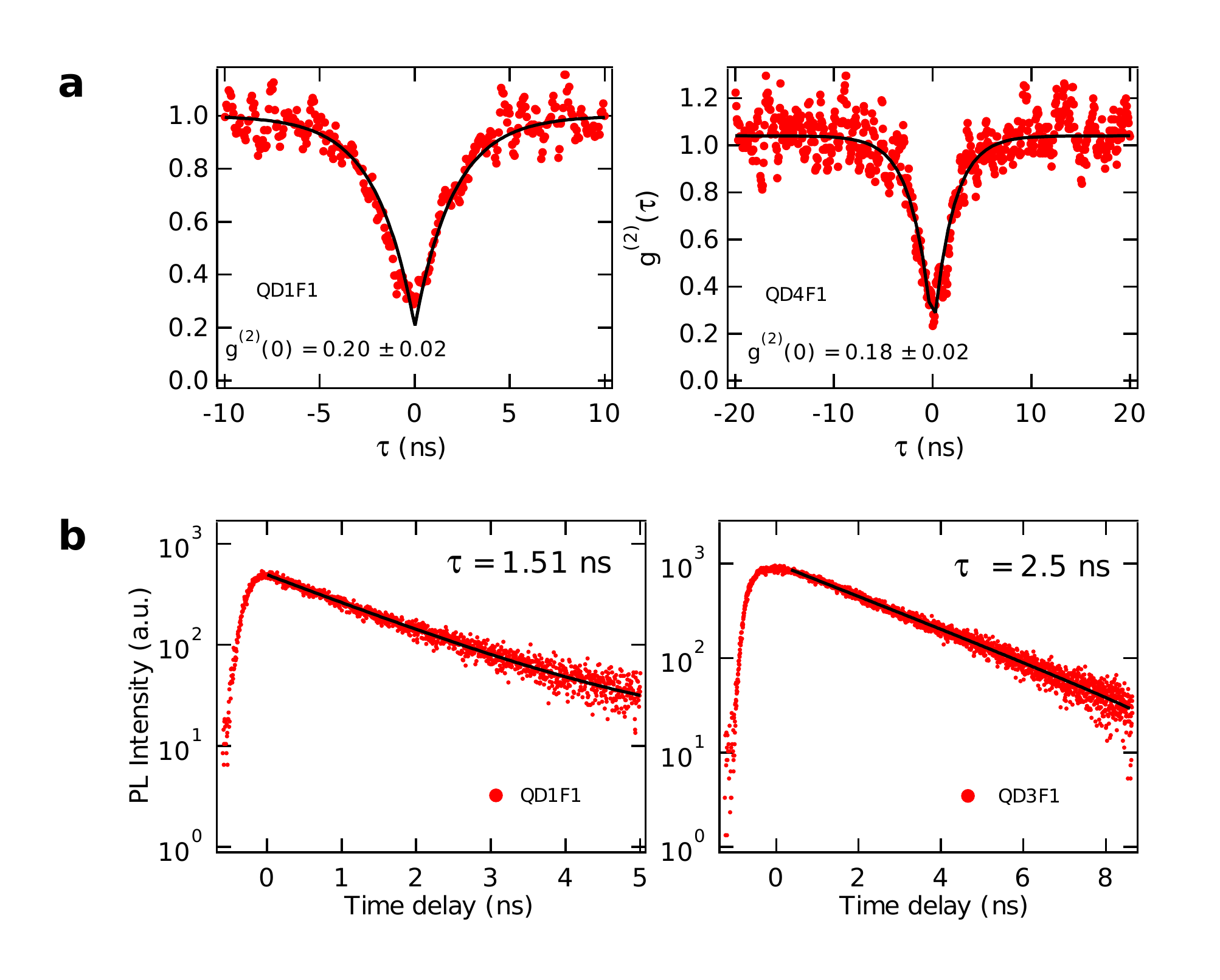}
\end{figure}

\clearpage

\begin{figure}
\includegraphics[scale=0.8]{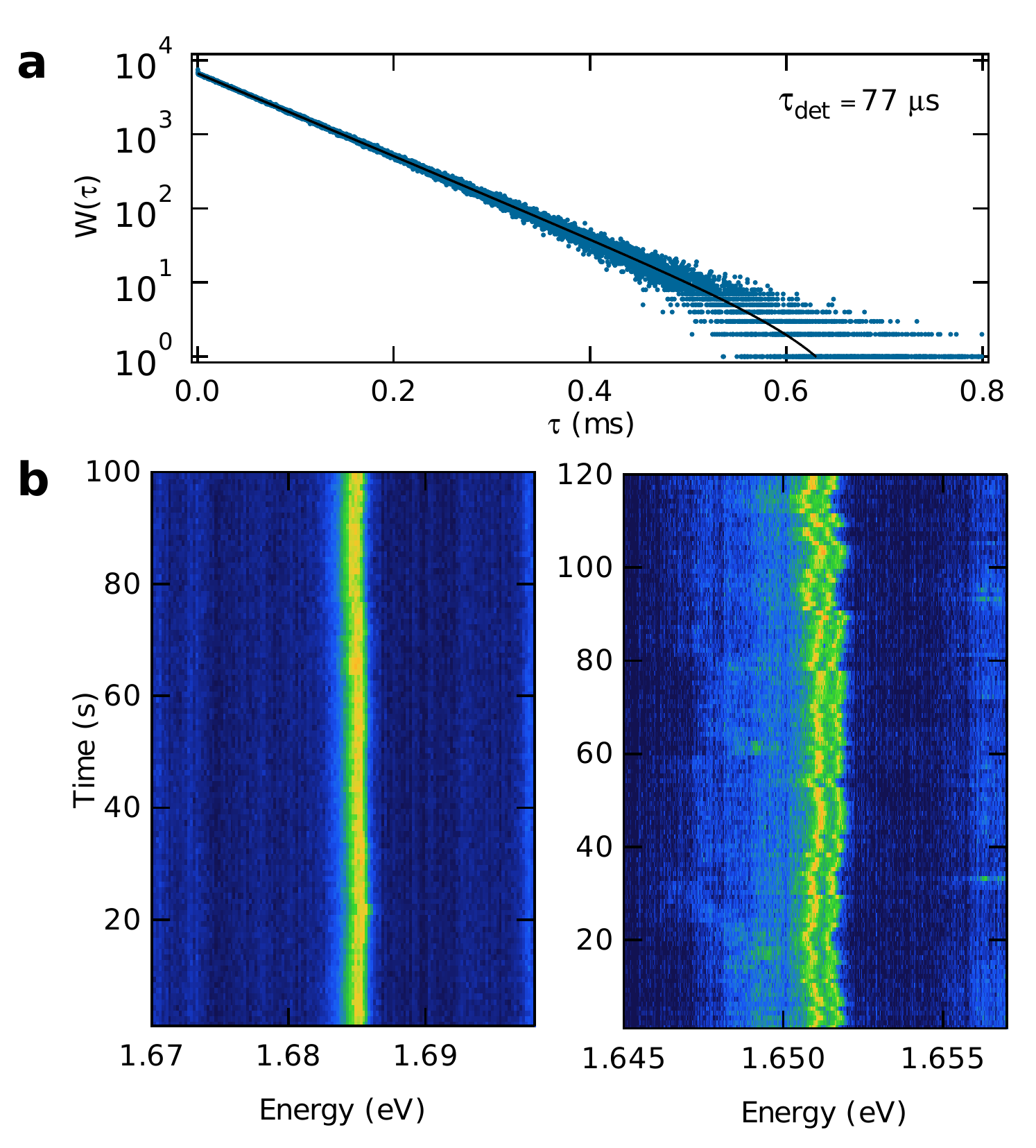}
\end{figure}

\clearpage

\begin{figure}
\includegraphics[scale=0.8]{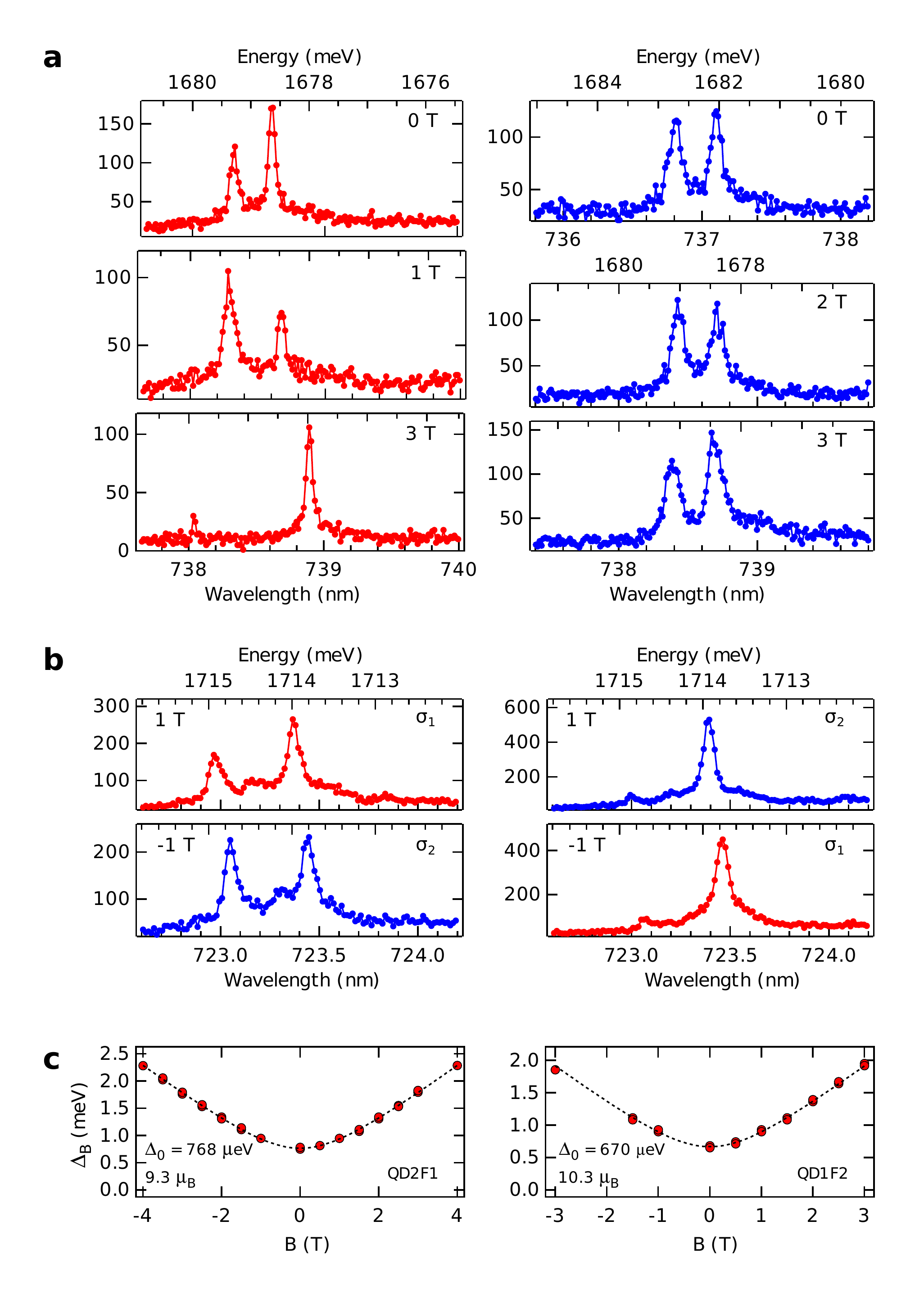}
\end{figure}

\clearpage

\textbf{Supplementary Information:}
\title{Optically Active Quantum Dots in Monolayer WSe$_2$}
\author{Ajit Srivastava$^1$}
\author{Meinrad Sidler$^1$}
\author{Adrien V. Allain$^2$}
\author{Dominik S. Lembke$^2$}
\author{Andras Kis$^2$}
\author{A. Imamo\u{g}lu$^1$}
\affiliation{$^1$Institute of Quantum Electronics, ETH Zurich, CH-8093
Zurich, Switzerland}
\affiliation{$^2$Electrical Engineering Institute, Ecole Polytechnique Federale de Lausanne (EPFL), CH-1015
Zurich, Switzerland.}

\maketitle

{\bf Figure S1}: (Left panel) Normalized photoluminescence spectra of flake0 for different incident laser powers. The different PL traces for exciton and trion
emission overlap with each other after normalizing with the incident
laser power implying a linear dependence of PL on incident laser
power. No broadening of the exciton and trion peak is observed even for the highest incident power of $\sim$ 320 $\mu$W. (Right panel) The impurity band emission, normalised with the incident laser power, exhibits saturation (sub-linear dependence) and drastic changes in the spectral features with increasing laser power.
\\

{\bf Figure S2}: Photoluminescence excitation (PLE) spectroscopy of quantum dots QD3F1 and QD1F2 performed by scanning a continuous wave tunable Ti:Sapphire laser near the free exciton resonance. Total integrated photoluminescence (PL) counts are plotted against laser wavelength showing clear enhancement of PL intensity at free exciton resonance (708 nm) for QD3F1. QD1F2 shows a similar enhancement however, at a slightly red-shifted wavelength. 
\\

{\bf Figure S3}: Magnetic field dependent splittings in the Faraday geometry for a few different dots with hyperbolic fits (see main text) to extract the magnetic moment and the exchange splitting at zero magnetic field.
\\

\clearpage

\begin{figure}
\includegraphics[scale=0.8]{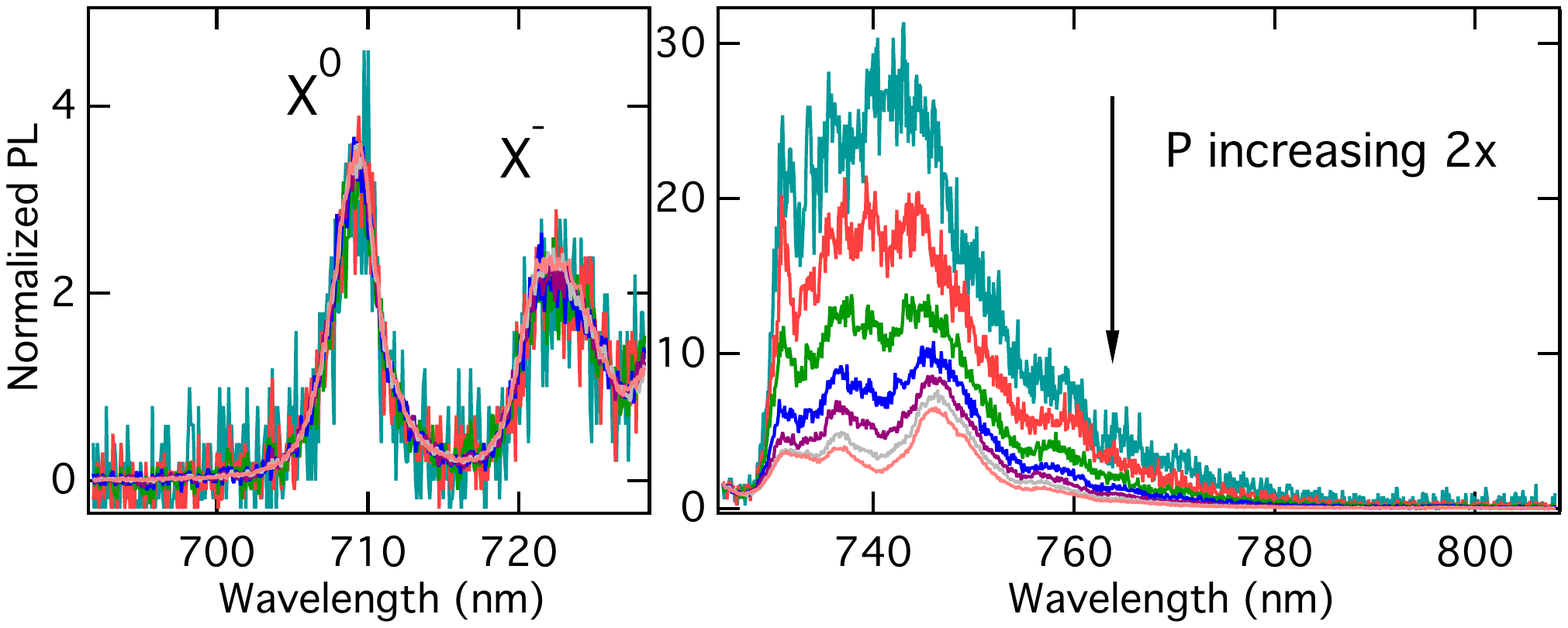}
\end{figure}

\clearpage

\begin{figure}
\includegraphics[scale=1]{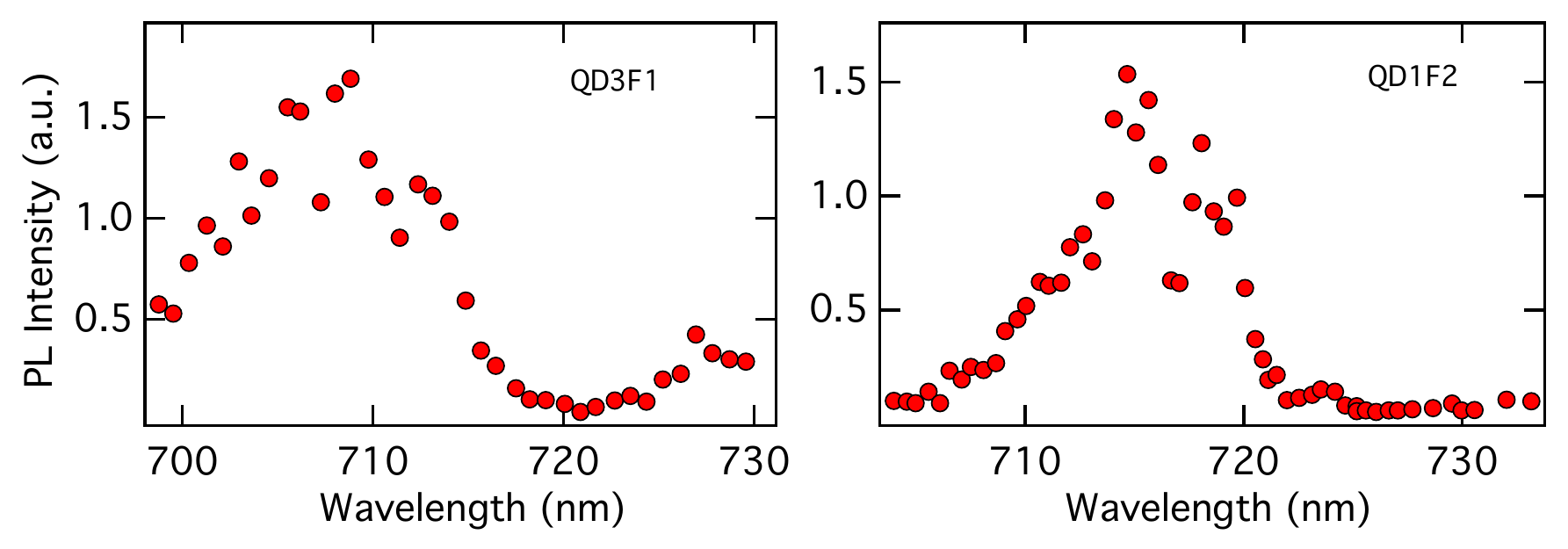}
\end{figure}

\begin{figure}
\includegraphics[scale=0.8]{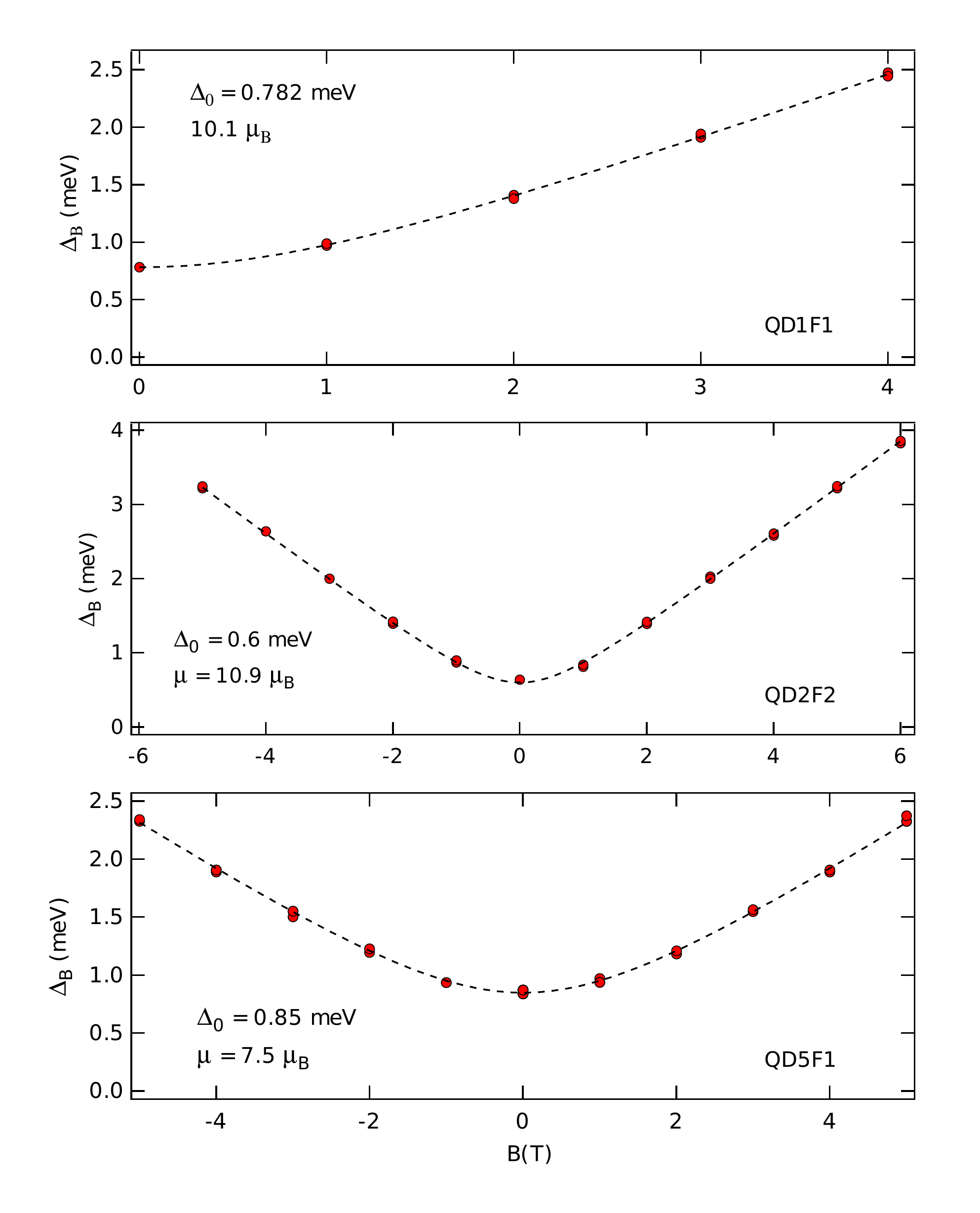}
\end{figure}

\end{document}